\documentclass[twocolumn,showpacs,preprintnumbers,amssymb]{revtex4}
\usepackage{graphicx, epsfig, amssymb} %include figure files
\usepackage{amsmath, amsfonts}
\usepackage{bm} 

\def\be{\begin{equation}}
\def\ee{\end{equation}}
\def\beq{\begin{eqnarray}}
\def\eeq{\end{eqnarray}}

\begin{document}

\centerline{}
%\vskip 1cm
\title{Black holes die hard: \\ can one spin-up a black hole past extremality?}

\author{
Mariam Bouhmadi-L\'opez,$^{1}$
\footnote{Electronic address: mariam.bouhmadi@ist.utl.pt}
Vitor Cardoso,$^{1,2}$
\footnote{Electronic address: vitor.cardoso@ist.utl.pt}
Andrea Nerozzi,$^{1}$
\footnote{Electronic address: andrea.nerozzi@ist.utl.pt}
Jorge V. Rocha,$^{1}$
\footnote{Electronic address: jorge.v.rocha@ist.utl.pt}
}
\affiliation{${^1}$ CENTRA,~Dept.~de F\'{\i}sica,~Instituto~Superior~T\'ecnico, Av.~Rovisco Pais 1, 1049 Lisboa, Portugal}
\affiliation{${^2}$ Department of Physics and Astronomy, The University of
Mississippi, University, MS 38677, USA}

\date{\today}

\begin{abstract}
A possible process to destroy a black hole consists on throwing point particles with sufficiently large angular momentum
into the black hole. In the case of Kerr black holes, it was shown by Wald that particles with dangerously large angular momentum are simply not captured by the hole, and thus the event horizon is not destroyed.
Here we reconsider this gedanken experiment for a variety of black hole geometries, from black holes in higher dimensions to black rings. We show that this particular way of destroying a black hole does not succeed and that Cosmic Censorship is preserved.
\end{abstract}

\pacs{04.70.Bw, 04.20.Dw}

\maketitle

%\tableofcontents

%\clearpage

%%%%%%%%%%%%%%%%%%%%%%%%%%%%%%%%%%%%%%%%%%%%%%%%%%%%%%%%%%%%%%%%%%%%%%%%%%%%%%%%
\section{Introduction}
%%%%%%%%%%%%%%%%%%%%%%%%%%%%%%%%%%%%%%%%%%%%%%%%%%%%%%%%%%%%%%%%%%%%%%%%%%%%%%%%
Black holes can be formed through the collapse of matter, through sufficiently high-energy
collisions of particles or quantum fluctuations in the early universe. Basically any process capable of confining a large portion of matter in a small enough space. Once formed, black holes are hard to kill. Quantum processes aside,
no known classical mechanism can destroy a black hole. One of such processes was considered by Wald 
\cite{wald} many years ago and revisited recently \cite{Matsas:2007bj,Matsas:2009ww,Hod:2008zza,Hubeny:1998ga,Jacobson:2009kt}.
It consists in throwing a point particle at a (four-dimensional) Kerr black hole with just the right angular momentum to spin the black hole up in such a way that eventually the horizon is disrupted. Indeed, the angular momentum of Kerr black holes is bounded by
$J\leq M^2$, thus {\it if it were possible} for the black hole to capture particles of high enough angular momenta, then one might exceed this bound,
possibly creating a naked singularity. Wald showed this cannot happen, as the potentially dangerous particles (i.e., those with large enough angular momentum) are never captured by the black hole \cite{wald}. 

The purpose of this short letter is to extend Wald's analysis to other spacetimes, in particular the Myers-Perry family of rotating black holes in higher dimensions \cite{Myers:1986un} and a large class of black rings in five dimensions \cite{Emparan:2001wn,Emparan:2004wy}.
This analysis is interesting because it allows one to test Cosmic Censorship in a very simple, yet realistic scenario.
The four-dimensional result indicates that no point particle thrown into a Kerr black hole can overcome the Kerr bound.
The analogous process for the case of {\it equal-mass} black holes was studied recently.
In Ref.~\cite{Sperhake:2009jz} the authors studied the collision at close to the speed of light of two equal-mass black holes with an arbitrary impact parameter. The end product of such collision was invariably a Kerr black hole, rotating at close to the maximum possible rate for certain critical impact parameters. No naked singularity was formed. Likewise, it might well be that the outcome of throwing point particles at black holes in other scenarios, for instance higher dimensions, provides some hints at what will happen in the full non-linear case. Thus, results obtained with ``point-particles'' could be used to understand numerical results in four- and even the on-going efforts in higher dimensions \cite{Berti:2010ce}.

The plan of the paper is as follows. In section \ref{sec:hdbh}, we review rotating black holes of spherical horizon topology in general D-dimensional space-time. Then we obtain the metric along the equatorial plane and consider the cases with a single rotation plane or with all angular momentum equal. In section \ref{potentialMP},  we obtain the effective potentials that describes the motion of a point-like particle along the equatorial plane in Myers-Perry (MP) geometry. We then study, in section \ref{sec:bhs}, how the dimensionless spin of a MP black hole evolves when it captures point particles. The analogous situation for neutral and dipole black rings in five dimensions is considered in section \ref{SUBH}. We conclude with some thoughts on possible extensions of our results.
%%%%%%%%%%%%%%%%%%%%%%%%%%%%%%%%%%%%%%%%%%%%%%%%%%%%%%%%%%%%%%%%%
\section{Higher dimensional black holes}
\label{sec:hdbh}
%%%%%%%%%%%%%%%%%%%%%%%%%%%%%%%%%%%%%%%%%%%%%%%%%%%%%%%%%%%%%%%%%
The geometries we are mainly concerned with describe rotating black holes in general 
$D-$dimensional spacetimes. In four dimensions, there is only one possible rotation axis for a
cylindrically symmetric spacetime, and there is therefore only one
angular momentum parameter.  In higher dimensions there are several
choices of rotation axis and there is a multitude of angular momentum
parameters, each referring to a particular rotation plane \cite{Myers:1986un}.  
The solution is described by a slightly different form depending on whether
the space-time dimension is even or odd. We briefly summarize the main results in the following.
Additional details can be found in the original work \cite{Myers:1986un}
(see also \cite{Nozawa:2005eu} where this discussion is taken from).
We use geometrical units with $G=c=1$.
%%%%%%%%%%%%%%%%%%%%%%%%%%%%%%%%%%%%%%%%%%%%%%%%%%%%%%%%%%%%%%%%%%%%%
\subsection{Even dimensions {\rm (}$D=2(d+1)${\rm )}}
%%%%%%%%%%%%%%%%%%%%%%%%%%%%%%%%%%%%%%%%%%%%%%%%%%%%%%%%%%%%%%%%%%%%%
The metric in even dimensions is given by
\begin{align}
\label{eq:mpeven}
 ds^2 &= - dt^2+r^2 d\alpha ^2+\sum_{i=1}^d 
\left(r^2+a_{i}^2\right)\left(d\mu _{i}^2
+\mu_{i}^2d\phi  _{i}^2\right)\nonumber 
\\
\qquad &+
\frac{M r}{{\it \Pi}{\it F}}\left(dt-\sum_{i=1}^d a_i \mu_{i}^2 d\phi _{i}
\right)^2
+\frac{{\it \Pi} F}{{\it \Pi}-M r}dr^2,
\end{align}
where
\be
F = 1-\sum_{i=1}^d \frac{a_i ^2 \mu _i ^2}{r^2+a_i^2} 
\,,\quad {\it \Pi}=\prod_{i=1}^d (r^2+a_i^2)\label{eq:F}\,,
\ee
and $\sum_{i=1}^d \mu _i ^2 +\alpha^2 = 1$, with $d\equiv D/2-1$.
The parameters $M$ and $a_i$
are related to the mass ${\cal M}$ and angular momenta ${\cal J}_i$
as
\beq
 && {\cal M}=\frac{D-2}{16\pi }A_{(D-2)}M,\\
 && {\cal J}_i =\frac{1}{8\pi }A_{(D-2)}M a_i ~~~~~~\, (i=1, \cdots, d)\,,
\eeq
where $A_{(D-2)}$ is the area of a unit ($D-2$)-sphere, which is given by
\beq
 A_{(D-2)}=\frac{2\pi ^{(D-1)/2}}{\Gamma ((D-1)/2)}\,.
\eeq

The event horizon is located at the zeroes of 
\beq
g^{rr}={{\it \Pi} -Mr\over {\it \Pi} F}\,.
\eeq
If at least one rotation parameter is set to zero, for example $a_1=0$, 
the equation for  the horizon is given by
\beq
{\it \Pi}-Mr=r^2 \left(\prod_{i\geq 2}^d (r^2+a_i^2)-\frac{M}{r}\right)=0\,.
\label{eq:horizon}
\eeq
In the case of $d\geq 2$,  i.e.  $D\geq 6$, Eq.~(\ref{eq:horizon}) always has a positive 
root, independently of the magnitude of $a_i$. We then find a regular black hole solution 
albeit with arbitrarily large angular momenta. This is one of the typical features of higher
dimensional black holes.
%%%%%%%%%%%%%%%%%%%%%%%%%%%%%%%%%%%%%%%%%%%%%%%%%%%%%%%%%%%%%%%%%%%%%%%
\subsection{Odd dimensions {\rm (}$D=2d+1${\rm )}}
%%%%%%%%%%%%%%%%%%%%%%%%%%%%%%%%%%%%%%%%%%%%%%%%%%%%%%%%%%%%%%%%%%%%%%%
In odd dimensions, the metric of a rotating black hole is slightly
changed from Eq. (\ref{eq:mpeven}). It is now given by
\begin{align}
\label{eq:mpodd}
ds^2 &= - dt^2+\sum_{i=1}^d\left(r^2+a_{i}^2\right)
\left(d\mu _{i}^2+\mu _{i}^2 d\phi _{i}^2\right) \nonumber\\
\qquad &+
\frac{M r^2}{{\it \Pi}{\it F}}\left(dt-\sum _{i=1}^d a_i \mu_{i}^2 d\phi _{i}\right)^2
+\frac{{\it \Pi F}}{{\it \Pi}-M r^2}dr^2 ,
\end{align}
with $\sum _{i=1}^d \mu _i ^2=1$. The definitions of $\it{\Pi}$ and $F$ remain
the same as in even dimensions while $d=(D-1)/2$.
We also find that if at least two angular momenta are set to zero, 
the remaining angular momenta can be arbitrarily large for $d\geq 3$,  i.e.  $D\geq 7$
as in the case of even dimensions.
%%%%%%%%%%%%%%%%%%%%%%%%%%%%%%%%%%%%%%%%%%%%%%%%%%%%%%%%%%%%%%%%%%%%%%%
\subsection{The five-dimensional rotating black hole}
%%%%%%%%%%%%%%%%%%%%%%%%%%%%%%%%%%%%%%%%%%%%%%%%%%%%%%%%%%%%%%%%%%%%%%%
The five-dimensional black hole is exceptional, because there is 
an upper bound for the angular momenta.
In Boyer-Lindquist coordinates, we can write down the five-dimensional 
black hole solution with two rotation parameters $a$ and $b$ as
\begin{align}
\label{eq:mp5}
ds^2 =& -dt^2 +\frac{\rho^2 r^2}{\Delta}dr^2+\rho^2 d\theta  ^2 \nonumber \\
      & +\frac{M}{\rho^2}(dt-a\sin^2 \theta d\varphi -b\cos^2 \theta d\psi)^2  \\
      & +(r^2+a^2)\sin^2 \theta d\varphi^2+(r^2+b^2)\cos^2 \theta d\psi^2 ,\nonumber
\end{align}
where
\beq
&& \rho^2=r^2+a^2\cos^2\theta +b^2\sin^2\theta ,\\
&& \Delta =(r^2+a^2)(r^2+b^2)-M r^2 .
\eeq
This can be obtained from~(\ref{eq:mpodd}) by setting
\beq
\mu_1 &=& \sin\theta\,,  \qquad  \phi_1=\varphi\,,  \qquad  a_1=a\,,\nonumber\\
\mu_2 &=& \cos\theta\,,  \qquad  \phi_2=\psi\,,     \qquad  a_2=b\,.
\eeq
The horizon appears where $\Delta =0$, which gives the location of 
the horizons, i.e.
\beq
r_{\pm}^2 &\equiv& \frac{M -(a^2 +b^2)}{2} \nonumber \\
          && \pm \frac{1}{2}\sqrt{[M -(a+b)^2][M-(a-b)^2]}\,.
\eeq
A sign change of rotation parameters $a, b$ 
simply reverses the direction of rotation.
The condition for the existence of an event horizon is
\be
\label{eq:above5mp}
M \geq (|a|+|b|)^2.
\ee
The outer and inner horizons coincide when $M =(|a|+|b|)^2$. 
The area of the event horizon is given by 
\be
{\cal A}_{H}={2\pi ^2 \over r_{+}}(r_+^2+a^2)(r_+^2+b^2).
\ee
The horizon vanishes if one of the spin parameters is set to zero
and the other approaches the extreme value  (e.g. $b=0$ and $a^2 \to M$), 
which corresponds to the appearance of a naked singularity.
When $(|a|+|b|)^2 \rightarrow M $ with $a\neq0$ and $b\neq0$, 
this corresponds to the extremal black hole with non-zero surface 
area and vanishing temperature.
%%%%%%%%%%%%%%%%%%%%%%%%%%%%%%%%%%%%%%%%%%%%%%%%%%%%%%%%%%%%%%%%%%%%%%%
\subsection{The metric along the equatorial plane}
%%%%%%%%%%%%%%%%%%%%%%%%%%%%%%%%%%%%%%%%%%%%%%%%%%%%%%%%%%%%%%%%%%%%%%%
We focus exclusively on the intuitively most dangerous process: particles falling in along the equator.
In this case, the metric and equations of motion simplify considerably. We will also 
consider two special sub-cases of the geometries discussed so far, (i) black holes with a 
single rotation parameter and (ii) black holes with all rotation parameter equal.
For simplicity, we will only discuss the motion of point particles in the ``equatorial plane'', which we now turn to.

The coordinates $\mu _i$ and $\alpha$ in the metric~(\ref{eq:mpeven})
are written explicitly by colatitude angles $\theta_i$ as follows:
\beq
 \left\{
\begin{array}{ll}
& \mu_1=\sin \theta _1\\
& \mu_2=\cos \theta_1\sin \theta_2  \\
& \quad \vdots  \\
& \mu_d=\cos \theta _1\cos \theta _2 \cdots \sin \theta_d  \\
& \alpha =\cos \theta _1 \cos \theta _2 \cdots \cos \theta_d\,.
\end{array}
\right.
\eeq
For the case of odd dimensionality the coordinate $\mu_d$ plays the role
of $\alpha$ and the above expression changes accordingly.
We then suppose that the orbits of particles are constrained on the
``equatorial'' plane $\theta_1=\theta_2=\cdots=\theta_d=\pi /2$.
Note however that since each coordinate $\mu_i$ is on equal footing, 
we can exchange the numbering of $\mu_i$, and find $d$ ``equatorial'' planes,
on which the orbits of particles are confined.

%%%%%%%%%%%%%%%%%%%%%%%%%%%%%%%%%%%%%%%%%%%%%%%%%%%%%%%%%%%%%%%%%%%%%%%
\subsection{A single rotation plane}
%%%%%%%%%%%%%%%%%%%%%%%%%%%%%%%%%%%%%%%%%%%%%%%%%%%%%%%%%%%%%%%%%%%%%%%
In this case, the metric along the equator is the same for even or odd $D$ and is given by
\beq
ds^2 &=& -{\Delta_D-a^2\over r^2}dt^2
    -{2a(r^2+a^2-\Delta_D) \over r^2}dtd\varphi \nonumber\\
     && +{(r^2+a^2)^2-\Delta_D a^2\over r^2} d\varphi^2
    +{r^2\over\Delta_D}dr^2\,,
\label{metric}
\eeq
where
\be
\Delta_D=r^2+a^2-M r^{5-D}\,.
\ee
For $D=4$, we recover the Kerr metric along the Equator.
The horizon is located at the zeroes of $\Delta_D$.

%%%%%%%%%%%%%%%%%%%%%%%%%%%%%%%%%%%%%%%%%%%%%%%%%%%%%%%%%%%%%%%%%%%%%%%
\subsection{All angular momenta equal}
%%%%%%%%%%%%%%%%%%%%%%%%%%%%%%%%%%%%%%%%%%%%%%%%%%%%%%%%%%%%%%%%%%%%%%%
The other extreme is when $a_i=a$ for all $i$.
In this case, we get
\beq
ds^2 &=& -\left(1-\frac{Mr}{f}\right)dt^2
   -2\frac{aMr}{f} dtd\varphi \nonumber\\
     &+& \left(r^2+a^2+\frac{a^2Mr}{f}\right) d\varphi^2
        +\frac{f}{(r^2+a^2)^{d}-Mr}dr^2\,,\nonumber
\eeq
for even $D$ and
\beq
ds^2 &=& -\left(1-\frac{Mr^2}{f}\right)dt^2
   -2\frac{aMr^2}{f} dtd\varphi \nonumber\\
     &+& \left(r^2+a^2+\frac{a^2Mr^2}{f}\right) d\varphi^2
   +\frac{f}{(r^2+a^2)^{d}-Mr^2}dr^2\,,\nonumber
\eeq
for odd $D$, with $f\equiv r^2(r^2+a^2)^{d-1}$.
The horizon, for odd $D$, is located at the zeroes of $(r^2+a^2)^{d}-Mr^2$.
In this case we find that the horizon radius and rotation parameters are limited as
\beq
r_+&\geq&{a\over\sqrt{d-1}}\,,\\
a&\leq&\left({(d-1)^{d-1}\over d^d }\right)^{1/(2(d-1))}M^{1/(2(d-1))}\,.
\label{extremal}
\eeq
%
%%%%%%%%%%%%%%%%%%%%%%%%%%%%%%%%%%%%%%%%%%%%%%%%%%%%%%%%%%%%%%%%%%%%%
\section{Effective potential for radial motion}\label{potentialMP}
%%%%%%%%%%%%%%%%%%%%%%%%%%%%%%%%%%%%%%%%%%%%%%%%%%%%%%%%%%%%%%%%%%%%%
With the use of the effective ``2+1'' dimensional metric along the equatorial plane,
it is very simple to write down the geodesic equations.
The conserved energy and angular momentum (per unit test-particle mass $m_0$ in the case of time-like geodesics~\footnote{For massless particles, the quantities $E$ and $L$ may be regarded as the energy and angular momentum.}) associated to the time-like and rotational Killing vectors are defined by
\be
E \equiv -g_{\mu\nu} (\partial/\partial t)^\mu \dot x^\nu \,, \qquad 
L \equiv g_{\mu\nu} (\partial/\partial \psi)^\mu \dot x^\nu\,,
\label{MandJ}
\ee
where the dot indicates derivation with respect to proper time.
Equations~(\ref{MandJ}) can be inverted to express $\dot t$ and $\dot \psi$ as linear combinations of $E$ and $L$.
To determine the `radial' motion one simply uses $g_{\mu\nu}\dot x^\mu\dot x^\nu = -\delta_1$, where $\delta_1=1,0$ for timelike and null geodesics, respectively.
%%%%%%%%%%%%%%%%%%%%%%%%%%%%%%%%%%%%%%%%%%%%%%%%%%%%%%%%%%%%%%%%%%%%
\subsection{A single rotation plane}
%%%%%%%%%%%%%%%%%%%%%%%%%%%%%%%%%%%%%%%%%%%%%%%%%%%%%%%%%%%%%%%%%%%%
Equatorial motion in the geometry~(\ref{metric}) can be reduced to the following radial equation~\cite{Cardoso:2008bp}
\beq\!\!\! \dot{r}^2&=&V_r\,,\label{potmyersperry} \\
\!\!\!r^2V_r&=&
\left [r^2E^2+\frac{M}{r^{D-3}}(aE-L)^2+(a^2E^2-L^2)
-\delta_1 \Delta_D \right]\,.\nonumber \eeq
We also have
\beq \dot{\varphi}&=&\frac{1}{\Delta_D}\left [\frac{aM}{r^{D-3}}E
+\left(1-\frac{M}{r^{D-3}}\right )L\right
]\,,\label{dotvarphi}\\
\dot{t}&=&\frac{1}{\Delta_D}\left [\left(r^2+a^2+\frac{a^2M}
{r^{D-3}}\right )E-\frac{aM}{r^{D-3}}L\right
]\,.\label{dott}
\eeq
The radial motion is completely governed by the potential $V_r$. If there are turning points outside the event horizon, then a particle coming from infinity can not reach the event horizon. Thus, the analysis we want to make is to study the maximum value of $L$ for which there are either no turning points, or all of them lie inside the event horizon.
%%%%%%%%%%%%%%%%%%%%%%%%%%%%%%%%%%%%%%%%%%%%%%%%%%%%%%%%%%%%%%%%%%%%%%%
\subsection{All angular momenta equal}
%%%%%%%%%%%%%%%%%%%%%%%%%%%%%%%%%%%%%%%%%%%%%%%%%%%%%%%%%%%%%%%%%%%%%%%
Similar equations can be written when all angular momenta are equal. For instance, for even $D$ we find,
\beq
r^3V_r &=& M\left(r^2+a^2\right)^{(4-D)/2}\left[r^2\delta_1+(L-aE)^2\right]\nonumber\\
       &+& r\left[(r^2+a^2)(E^2-\delta_1)-L^2\right]\,,
\eeq
while for odd $D$ we obtain
\beq
r^2V_r &=& M\left(r^2+a^2\right)^{(3-D)/2}\left[r^2\delta_1+(L-aE)^2\right]\nonumber\\
       &+& \left[(r^2+a^2)(E^2-\delta_1)-L^2\right]\,.
\eeq
Specializing to the case of $D=5$ the above equation reduces to
\beq
r^2(r^2+a^2)V_r &=& Mr^2\delta_1+M(L-aE)^2 \\
                &+& (r^2+a^2)^2(E^2-\delta_1)-(r^2+a^2)L^2\,.\nonumber
\eeq
%

%%%%%%%%%%%%%%%%%%%%%%%%%%%%%%%%%%%%%%%%%%%%%%%%%%%%%%%%%%%%%%%%%%%%%%%%%%%%%%%%
\section{Spinning-up a black hole by throwing point particles\label{sec:bhs}}
%%%%%%%%%%%%%%%%%%%%%%%%%%%%%%%%%%%%%%%%%%%%%%%%%%%%%%%%%%%%%%%%%%%%%%%%%%%%%%%%
Let us try to spin-up a BH with mass ${\cal M}_0$ and angular momentum ${\cal J}_0$ in general $D$ spacetime dimensions.
For that, we throw in a particle of mass $m_0$ with angular momentum $\delta\,{\cal J}=m_0L$ and energy $\delta\,{\cal M}=m_0E$, such that $\delta{\cal M}\ll {\cal M}_0$ and $\delta{\cal J}\ll {\cal J}_0$.
Upon absorption of this particle, the dimensionless spin of the BH~\footnote{It is also common to normalize the dimensionless spin $j$ by a factor $\sqrt{27\pi/32}$, so that the extremal five-dimensional MP and the singular fat black ring have $j=1$.} 
\be
j\equiv \frac{{\cal J}}{{\cal M}^{\frac{D-2}{D-3}}}\,,
\ee
changes to
\be
j=j_0+\delta j\,,
\ee
where the subscript stands for initial parameters of the BH and 
\be
\delta j=\frac{m_0}{{\cal M}_0}\left(\frac{L}{{\cal M}_0^{\frac{1}{D-3}}}-Ej_0\frac{D-2}{D-3}\right)\,.\label{evspin}
\ee
%
%%%%%%%%%%%%%%%%%%%%%%%%%%%%%%%%%%%%%%%%%%%%%%%%%%%%%%%%%%%%%%%%%%%%%%%%%%%%%%%%%%%%%%%%
\subsection{Single rotation parameter}
%%%%%%%%%%%%%%%%%%%%%%%%%%%%%%%%%%%%%%%%%%%%%%%%%%%%%%%%%%%%%%%%%%%%%%%%%%%%%%%%%%%%%%%%
We start with the $D=5$ case, which is simple enough that it allows an explicit solution \cite{Cardoso:2008bp}.
Let's focus on co-rotating geodesics, since these are the only ones of significance here.
For capture to occur, we find that the angular momentum has to be smaller than the critical value
\be
L_{\rm crit}=E\sqrt{M}+\sqrt{E^2-1}\left(\sqrt{M}-a\right)\,.
\ee
For large $E$ it tends to $\frac{L_{\rm crit}}{E}\to 2\sqrt{M}-a$, which also corresponds
to the values of the null circular geodesic, as could be expected \cite{Cardoso:2008bp}.
Eq. (\ref{evspin}) yields
\be
(\delta\,j)_{\rm max}=\frac{m_0}{{\cal M}_0}\left(\frac{E\sqrt{M}+\sqrt{E^2-1}\left(\sqrt{M}-a\right)}{\sqrt{{\cal M}_0}}-\frac{3E}{2}j_0\right)\,.
\ee
For $D=5$, we also have
\be
{\cal M}=3\pi M/8\,, \qquad
{\cal J}=2{\cal M}a/3\,.
\label{5DMJ}
\ee
Thus we can write $j_0=2a/(3\sqrt{{\cal M}_0})$ and
\beq
(\delta\,j)_{\rm max} &=& \frac{m_0}{{\cal M}_0^{3/2}}\left(\sqrt{M}-a\right)\left(E+\sqrt{E^2-1}\right) \nonumber\\
  &=& \frac{4m_0}{\pi M} \left(\sqrt{\frac{32}{27\pi}}-j_0\right)\left(E+\sqrt{E^2-1}\right) \,.
\eeq

Therefore, for $a<\sqrt{M}$, or equivalently for $j_0^2<\frac{32}{27\pi}$, the BH can be spun-up by the capture of particles.
This spinning-up process ceases when the rotation reaches $a=\sqrt{M}$.  As in four dimensions, in $D=5$ we can also spin the BH to the extremal limit and not further than that~\cite{Elvang:2006dd}.

\begin{figure}[htpb!]
\includegraphics[width=9cm]{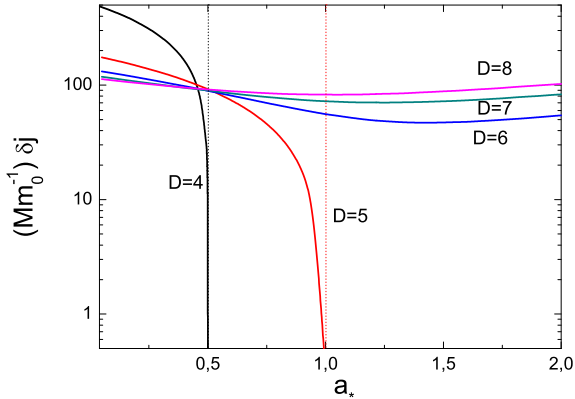}
\caption{\label{fig:oneJ} This figure shows the maximum increase in spin, $\frac{M}{m_0}(\delta j)_{\rm max}$ caused by a particle with $E/m_0=100$ falling into a Myers-Perry black hole with a single rotation parameter. The dimensionless rotation parameter $a_*$ is defined as $a_*\equiv \frac{a}{M^{1/(D-3)}}$.
Notice that it is not possible to spin-up an extremal black hole
(for $D=4,5$, the extremal value is marked with a dotted line).} 
\end{figure}

What about general $D$? Unfortunately, an exact analysis such as the previous one for $D=5$ does not seem to be possible. 
We have numerically searched for the critical angular momentum, and computed $\delta j$ in Eq. (\ref{evspin}). The results, which are summarized in Fig.~\ref{fig:oneJ}, are clear: neutral black holes in four and five spacetime dimensions with a single rotation cannot be spun-up past extremality.
For larger $D$, there is no extremal limit, and the black holes can be spun-up to an arbitrarily high rotation.

One can obtain analytic expressions in the limit that both the rotation of the hole and the energy of the incoming particle are large.
In this case, it is sufficient to focus attention on the (co-rotating) circular null geodesic with $r=r_c$ as the geodesic with maximum possible impact parameter that can still be captured. This geodesic has \cite{Cardoso:2008bp}
\be
\frac{L}{E}=a+\sqrt{\frac{2r_c^{D-1}}{(D-3)M}}\,,
\ee
and for large $a$ we get $\frac{L}{E}\sim a$ \cite{Cardoso:2008bp}. Thus, from equation (\ref{evspin}) we get 
\beq
\delta\,j&=&\frac{m_0}{{\cal M}_0}\left(\frac{Ea}{{\cal M}_0^{1/(D-3)}}-Ej_0\frac{D-2}{D-3}\right)\nonumber\\
&=&\frac{m_0(D-2)j_0E}{{\cal M}_0}\left(\frac{1}{2}-\frac{1}{D-3}\right)\,.
\eeq
In agreement with the numerical results, $\delta j$ is always positive, {\it in this limit}.

\begin{figure}[htpb!]
\includegraphics[width=9cm]{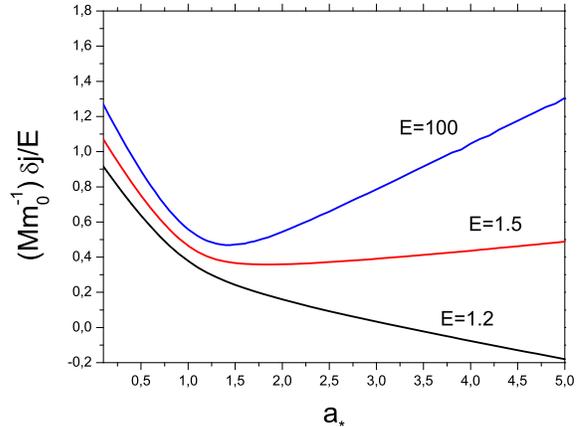}
\caption{\label{fig:D6Es} This figure shows the maximum increase in spin, $\frac{M}{m_0}(\delta j)_{\rm max}/E$ caused by a particle with energy $E=100,1.5,1.2$
falling into a Myers-Perry black hole with $D=6$ and for  a single rotation parameter. The dimensionless rotation parameter $a_*$ is defined as $a_*\equiv \frac{a}{M^{1/(D-3)}}$. For $E=1.2$, $\frac{M}{m_0}\delta j$ becomes negative at $a_*=3.3$.} 
\end{figure}

Our results also show that the variation in dimensionless spin depends sensitively on the energy of the point particle. For instance, Fig.~\ref{fig:D6Es} depicts how the spin of a $D=6$ Myers-Perry black holes depends on the energy of the captured particle.
We find a qualitative change in the behavior of $\delta j$ for low energy. More specifically, there is a critical energy $E_{\rm{crit}}$ above which the dimensionless spin parameter is a growing function of the dimensionless rotation parameter $a_*\equiv \frac{a}{M^{1/(D-3)}}$ (at large $a_*$), while for values of $E<E_{\rm{crit}}$, $\delta j$ is a decreasing function of $a_*\equiv \frac{a}{M^{1/(D-3)}}$. Indeed, in this case $\delta j$ eventually becomes negative. The value of $E_{\rm{crit}}$ depends on the spacetime dimension: 
\beq
D&=&6: \quad 1.34< E_{\rm{crit}}<1.35\,,\nonumber\\
D&=&7: \quad 1.15< E_{\rm{crit}}<1.16\,,\nonumber\\
D&=&8: \quad 1.09< E_{\rm{crit}}<1.10\,.\nonumber
\eeq
As can be noticed, $E_{\rm{crit}}$ gets smaller as the spacetime dimension increases.

%%%%%%%%%%%%%%%%%%%%%%%%%%%%%%%%%%%%%%%%%%%%%%%%%%%%%%%%%%%%%%%%%%%%%%%%%%%%%%%%%%%%%%%%
\subsection{All angular momenta equal}
%%%%%%%%%%%%%%%%%%%%%%%%%%%%%%%%%%%%%%%%%%%%%%%%%%%%%%%%%%%%%%%%%%%%%%%%%%%%%%%%%%%%%%%%
As for the singly-spinning case, the situation in which all angular momenta are equal can be solved analytically in $D=5$.
Again we focus on time-like co-rotating geodesics.
For capture to occur, we find that the angular momentum has to be smaller than the critical value
\be
L_{\rm crit}=E\sqrt{M}+\sqrt{(E^2-1)\left(M-2a\sqrt{M}\right)}\,.
\ee
For large $E$ it tends to $\frac{L_{\rm crit}}{E}\to \sqrt{M}+\sqrt{M-2a\sqrt{M}}$, which also corresponds
to the values of the null circular geodesic, as should be expected.
Eq.~(\ref{evspin}), together with eq.~(\ref{5DMJ}), yields
\beq
(\delta\,j)_{\rm max} &=& \frac{m_0}{{\cal M}_0^{3/2}}\left[ E\left(\sqrt{M}-a\right)\right. \nonumber\\
            &&  +\left.\sqrt{(E^2-1)\left(M-2a\sqrt{M}\right)} \right]\,.
\eeq

Notice that for $D=5$ the extremal value of the spin parameter is $a=\sqrt{M}/2$, as can be easily seen from eq.~(\ref{extremal}).
In this case, even when we take the extremal limit we obtain a positive maximum increment in the dimensionless spin of the BH:
\be
(\delta\,j)_{\rm max} \to \frac{m_0}{{\cal M}_0^{3/2}} \frac{E\sqrt{M}}{2}\,.
\ee
Nevertheless, this does not imply that the cosmic censorship conjecture is violated for 5$D$ Myers-Perry with both angular momenta equal.
We are changing the angular momentum ${\cal J}_1$ by throwing in a massive particle with angular momentum $\delta{\cal J}_1$.  The mass of the BH also increases by $\delta{\cal M}$ and we have shown that $\delta j_1$ can be positive.  However, in the process  $\delta j_2$ {\em decreases} and so we are left with a BH with different angular momenta for which the extremal bound~(\ref{extremal}) no longer applies. Instead it is replaced by
\be
|j_1|+|j_2| \leq \sqrt{\frac{32}{27\pi}}  \quad \Longleftrightarrow \quad
|a_1|+|a_2| \leq \sqrt{M}\,.
\ee
In fact, from eq.~(\ref{evspin}) we find $\delta j_2 = - m_0 {\cal M}_0^{-3/2}Ea$,
so that
\beq
\delta j_1 + \delta j_2 &=& \frac{m_0}{{\cal M}_0^{3/2}}\left[ E\left(\sqrt{M}-2a\right)\right. \nonumber\\
          &&  +\left.\sqrt{(E^2-1)\left(M-2a\sqrt{M}\right)} \right]\,.
\eeq
This shows that, taking the extremal limit, the change in angular momenta produced by throwing one test particle into a 5D Myers-Perry BH with two equal angular momenta still yields an extremal configuration, albeit with different spin parameters.

The study just performed is hard to generalize to higher dimensions because the surface of extremal solutions becomes more complicated as the dimension increases.  However, one can avoid this by the following trick: instead of throwing in one test particle, consider $d$ particles following similar geodesics along the $d$ orthogonal rotation planes.  The final black hole will also have all angular momenta equal. For the $5D$ case it is easy to reproduce this situation -- the only difference relative to the previous calculation is that, since we are throwing in {\em two} particles the increment in mass is doubled. Therefore,
\beq
(\delta\,j_1)_{\rm max} &=& (\delta\,j_2)_{\rm max} =  
\frac{m_0}{{\cal M}_0^{3/2}}\left[ E\left(\sqrt{M}-2a\right)\right. \nonumber\\
            &&  +\left.\sqrt{(E^2-1)\left(M-2a\sqrt{M}\right)} \right]\,,
\eeq
which vanishes in the extremal limit, $a \to \sqrt{M}/2$.

\begin{figure*}[htb]
\begin{center}
\begin{tabular}{cc}
\includegraphics[scale=0.3,clip=true,angle=0]{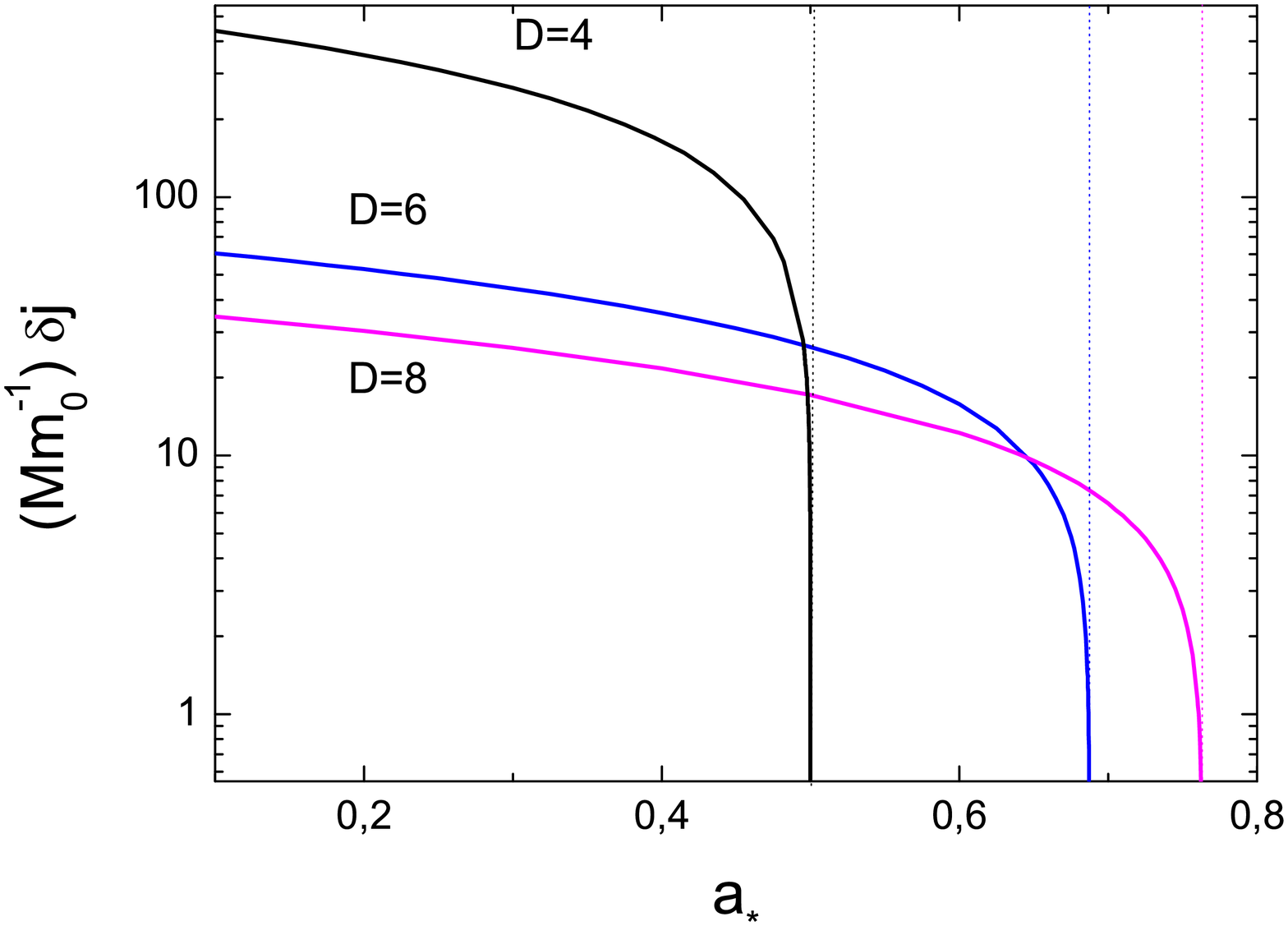}
&\includegraphics[scale=0.3,clip=true,angle=0]{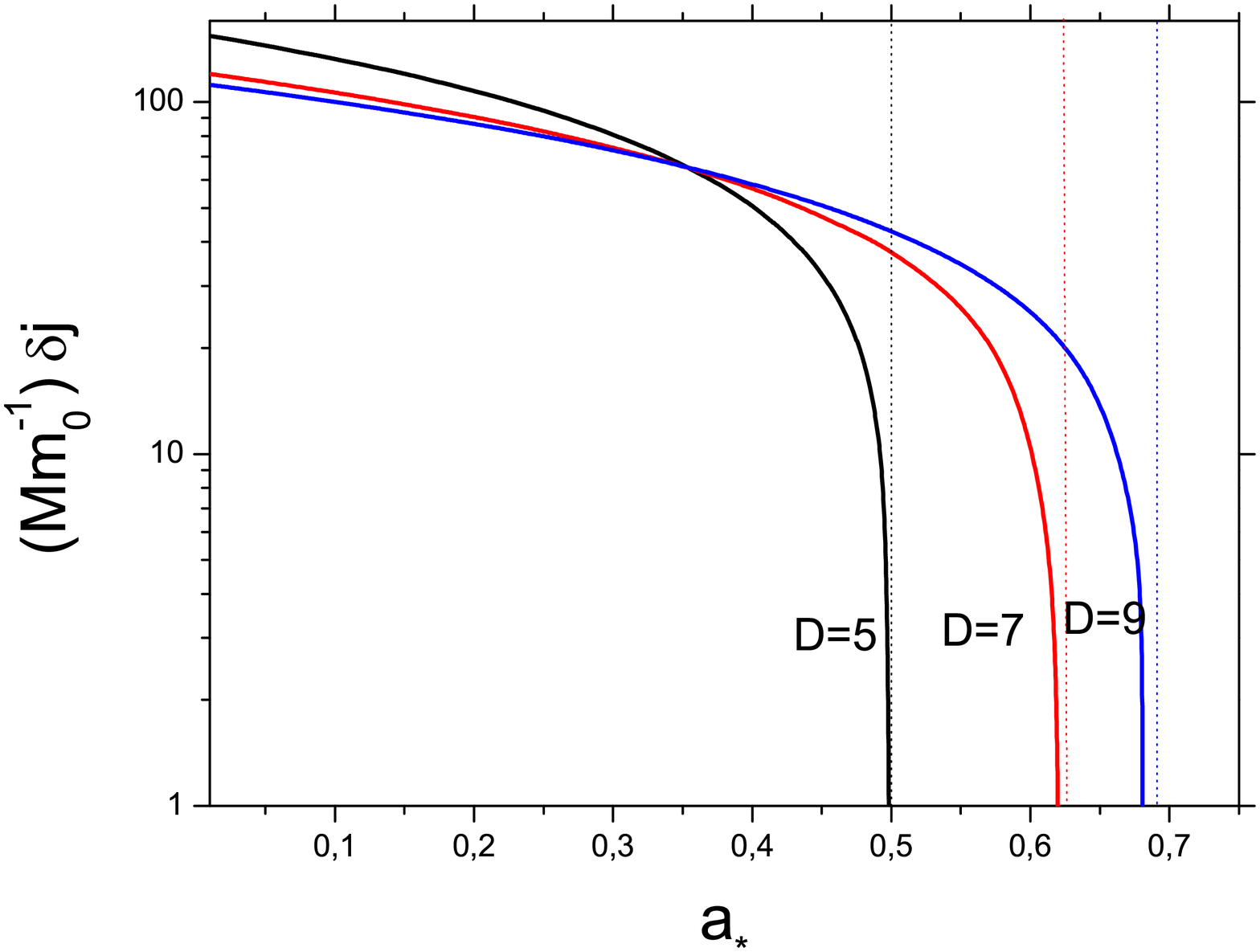}
\end{tabular}
\end{center}
\caption{\label{fig:EqualJ} This figure shows the maximum increase in spin, $\frac{M}{m_0}(\delta j)_{\rm max}$ caused by $d$ particles with $E/m_0=100$
falling into a Myers-Perry black hole with all rotation parameters equal. The left panel considers even $D$, the right panel refers to odd spacetime dimensions $D$. It is not possible to spin-up an extremal black hole, marked with a dotted line. The dimensionless rotation parameter $a_*\equiv \frac{a}{M^{1/(D-3)}}$.} \end{figure*}

The same {\it rationale} can be applied for higher dimensions; i.e. 
\be
\delta j=\frac{m_0}{{\cal M}_0}\left(\frac{L}{{\cal M}_0^{\frac{1}{D-3}}}-dEj_0\frac{D-2}{D-3}\right)\,,
\ee
and the results are presented in Fig.~\ref{fig:EqualJ}. In full analogy with the singly spinning case in D=5,6 in which the spin is bounded, we cannot exceed the extremal limit by throwing in test particles.

%
%%%%%%%%%%%%%%%%%%%%%%%%%%%%%%%%%%%%%%%%%%%%%%%%%%%%%%%%%%%%%%%%%%%%%%%
\section{Spinning-up black rings}\label{SUBH}
%%%%%%%%%%%%%%%%%%%%%%%%%%%%%%%%%%%%%%%%%%%%%%%%%%%%%%%%%%%%%%%%%%%%%%%
In this section we consider the case of singly spinning black rings~(see~\cite{Emparan:2006mm} for a review). The neutral black ring was obtained by Emparan and Reall~\cite{Emparan:2001wn} and is a solution of vacuum gravity in five dimensions featuring a horizon with spatial topology $S^1 \times S^2$. We shall also consider the more general case of the dipole black ring discovered in~\cite{Emparan:2004wy} (however, note that we restrict to the non-dilatonic solution).

It is well known that the neutral ring has no upper bound on its dimensionless angular momentum $j$. However, there are two families of black rings that coexist in a certain range of parameters, the `fat' and the `thin' rings, and for the fat ring branch there is an upper bound on $j$. Ref.~\cite{Elvang:2006dd} has shown that it is not possible to overspin a fat ring with massless particles. We first extend this result, in Section~\ref{sec:neutrings}, to the case of absorption of a massive particle by the neutral black ring.

Then we consider the case of the dipole ring in Section~\ref{sec:diprings}. For our purposes the main novel feature of the dipole ring with respect to its neutral counterpart is that it possesses both lower and {\em upper} bounds on the spin. In this case the distinction between `fat' and `thin' rings still holds, but it is determined by the dipole charge parameter.

Let us first collect here the necessary results. The metric can be expressed in the following form~\cite{Emparan:2004wy}:
\begin{flalign}
&ds^2 = -\frac{F(y)}{F(x)}\frac{H(x)}{H(y)} \left( dt-C R \frac{1+y}{F(y)}d\psi \right)^2 \nonumber \\
& +\frac{R^2 F(x)H(x)H^2(y)}{(x-y)^2} \left[ -\frac{G(y)}{F(y)H^3(y)}d\psi^2 + \frac{G(x)}{F(x)H^3(x)}d\phi^2 \right] \nonumber \\
& +\frac{R^2 F(x)H(x)H^2(y)}{(x-y)^2} \left[ - \frac{dy^2}{G(y)} + \frac{dx^2}{G(x)}  \right], 
\label{metric:dipole}
\end{flalign}
where
\beq
F(\xi) &=& 1+\lambda \xi\,, \\
G(\xi) &=& (1-\xi^2)(1+\nu \xi)\,, \\
H(\xi) &=& 1-\mu\xi\,,
\eeq
and
\be
C=\sqrt{\lambda(\lambda-\nu)\frac{1+\lambda}{1-\lambda}}\,.
\ee

In general, the metric~(\ref{metric:dipole}) is plagued with conical singularities but these are absent when the parameters satisfy
\be
\frac{1-\lambda}{1+\lambda}\left(\frac{1+\mu}{1-\mu}\right)^3=\left(\frac{1-\nu}{1+\nu}\right)^2.
\label{balance}
\ee
This situation, which we shall assume from now on, corresponds to a balanced ring in the sense that the centrifugal force compensates for the tension and self-attraction of the ring.  Therefore, this solution has three free parameters, which can be taken to be $R$, $\nu$ and $\mu$, but the former has dimensions of length and drops out of all dimensionless ratios.  The parameter $R$ measures the radius of the ring, whereas $R\nu$ can be viewed as the radius of the $S^2$ at the horizon. Finally $\mu$ is associated to the dipole charge. Setting $\mu=0$ one obtains the neutral black ring, for which the regularity condition~(\ref{balance}) becomes
\be
\lambda = \frac{2\nu}{1+\nu^2}\,.
\ee

Restricting to the neutral ring, the coordinate $y$ takes values in the interval $(-\infty,-1]$ whereas $x$ is restricted to $x\in[-1,1]$.  Surfaces of constant $y$ are ring-shaped.  The surface $y=-1$ is identified with the axis of rotation in the $\psi$ direction.  The coordinate $x$ can be viewed as a polar coordinate on the $S^2$.  The axis of rotation along the angle $\phi$ corresponds to $x=\pm 1$. The $+$ sign yields the central disk bounded by the ring and the $-$ sign gives the $\phi$-axis outside the ring.  The dimensionless parameters $\nu$ and $\lambda$ take values in the range
\be
0<\nu \leq \lambda <1\,.
\ee
The outer horizon lies at $y=-1/\nu$ and there is an ergosurface at $y=-1/\lambda$.  Finally, at $y=-\infty$ the solution reveals a space-like singularity.

In the presence of a finite dipole charge the coordinate $y$ can be extended across $|y|=\infty$ to the interval $(1/\mu,+\infty)$. The dipole parameter varies in the range
\be
0\leq\mu<1\,,
\ee
and a curvature singularity appears only as $y\rightarrow1/\mu^+$, while there is an inner horizon at $y=-\infty$. When $\nu\to0$ the outer and inner horizons become degenerate and this corresponds to the extremal limit~\cite{Emparan:2004wy}.

The mass and angular momentum of the singly spinning balanced dipole ring are given by
\beq
{\cal M}_0&=&\frac{3\pi R^2}{4}\frac{(1+\mu)^3}{1-\nu}\left(\lambda+\frac{\mu(1-\lambda)}{1+\mu}\right),
\label{eqn:dipolemass} \\
{\cal J}_0&=&\frac{\pi R^3}{2} \frac{(1+\mu)^{9/2}\sqrt{\lambda(\lambda-\nu)(1+\lambda)}}{(1-\nu)^2},
\label{eqn:dipoleJ}
\eeq
while the dipole charge is~\cite{Emparan:2004wy}
\be
{\cal Q}_0 = \sqrt{3}R \frac{(1+\mu)\sqrt{\mu(\mu+\nu)(1-\lambda)}}{(1-\nu)\sqrt{1-\mu}}\,.
\label{eqn:dipolecharge}
\ee
From the formulas for the mass and dipole charge given in Eqs.~(\ref{eqn:dipolemass}) and~(\ref{eqn:dipolecharge}) a dimensionless ratio can be obtained by
\be
q \equiv \frac{\cal Q}{{\cal M}^{1/2}}\,.
\label{eqn:dipcharge}
\ee

Geodesics in the background of a neutral black ring have been analyzed in reference~\cite{Hoskisson:2007zk}.
Here we generalize to the dipole ring but restrict our attention to geodesic motion in the equatorial plane outside the ring, i.e., $x=-1$. 
Thus, the $\phi$-angle drops out and we are left with the metric
\beq
ds^2 &=& -\frac{(1+\mu)F(y)}{(1-\lambda)H(y)}dt^2 + \frac{2R C(1+\mu)(1+y)}{(1-\lambda)H(y)} dt d\psi \nonumber\\
&& - \frac{R^2(1+\mu)}{H(y)F(y)} \left[ \frac{C^2}{1-\lambda} (1+y)^2 + \frac{(1-\lambda)G(y)}{(1+y)^2} \right]d\psi^2 \nonumber\\
&& - \frac{R^2 (1-\lambda)(1+\mu)H(y)^2}{(1+y)^2 G(y)} dy^2\,.
\label{metric:dipringequator}
\eeq
Equations~(\ref{MandJ}) can be inverted to yield
\begin{flalign}
&\dot t = \frac{H(y)}{(1+\mu)(1-\lambda)F(y)G(y)} \nonumber\\
&\times \left\{ \left[ (1-\lambda)^2G(y) + C^2(1+y)^4 \right] E - C(1+y)^3 F(y)\frac{L}{R} \right\}\,, \nonumber\\
&\dot \psi = \frac{H(y)(1+y)^2}{R(1+\mu)(1-\lambda)G(y)} \left[ C(1+y)E - F(y)\frac{L}{R} \right]\,,
\label{eqn:psidottdot}
\end{flalign}
and the radial motion is governed by the following equation:
%
%\begin{flalign}
%&\dot y^2 = V_y\,, \nonumber\\
%&R^2 V_y = -(1+y)^3 \left( \alpha_3 y^3 + \alpha_2 y^2 + \alpha_1 y + \alpha_0 \right)\,,
%\label{eqn:potdipole}
%\end{flalign}
%
%where
%
%\beq
%\alpha_3 &=&-\frac{\mu\mathcal{C}_2}{1+\mu}\,, \nonumber\\
%\alpha_2 &=&\frac{\mathcal{C}_2-\mu\mathcal{C}_1}{1+\mu}+\frac{\nu\delta_1}{1-\lambda}\,, \nonumber\\
%\alpha_1 &=&\frac{\mathcal{C}_1-\mu\mathcal{C}_0}{1+\mu}+\frac{(1-\nu)\delta_1}{1-\lambda}\,, \nonumber\\
%\alpha_0 &=&\frac{\mathcal{C}_0}{1+\mu}-\frac{\delta_1}{1-\lambda}\,, \nonumber
%\eeq
%
%and
%
%\beq
%\mathcal{C}_2 &=& \frac{1}{\lambda(1-\lambda)^2} \left( C E - \lambda\frac{L}{R} \right)^2\,, \nonumber\\
%\mathcal{C}_1 &=& \left[ \frac{C^2}{\lambda(1-\lambda)^2}\left(3-\frac{1}{\lambda}\right)-\frac{\nu}{\lambda} \right] E^2 
 %     -\frac{4C}{(1-\lambda)^2} E\frac{L}{R} \nonumber\\
  % && +\frac{1+\lambda}{(1-\lambda)^2}\left(\frac{L}{R}\right)^2\,, \nonumber\\
%\mathcal{C}_0 &=& \left[ \frac{C^2}{\lambda(1-\lambda)^2}\left(3-\frac{3}{\lambda}+\frac{1}{\lambda^2}\right) +\frac{\nu}{\lambda}\left(1+\frac{1}{\lambda}\right)-\frac{1}{\lambda} \right] E^2 \nonumber\\
 %  && -\frac{2C}{(1-\lambda)^2} E\frac{L}{R} + \frac{1}{(1-\lambda)^2}\left(\frac{L}{R}\right)^2\,.
%\eeq
%

%%%%%%%%%%%%%%%%%%%%%%%%%%%%%%%%%%%%%%%%%%%%%%%%%%%%%%%%%%%%%%%%%
%Here's an equivalent form for the radial potential equation:
%
\beq
\dot y^2 &=& V_y\,, \nonumber\\
R^2 V_y &=& \frac{(1+y)^3}{(1+\mu)H(y)^2} \left[ -\frac{H(y)P(y)}{(1+\mu)(1-\lambda)^3}E^2 \right. \nonumber\\
  &+& \frac{2C(1+y)^2H(y)}{(1+\mu)(1-\lambda)^2} E\frac{L}{R} - \frac{(1+y)F(y)H(y)}{(1+\mu)(1-\lambda)^2} \left(\frac{L}{R}\right)^2 \nonumber\\
  &+& \left. \frac{(1-y)(1+\nu y)}{1-\lambda}\delta_1 \right]\,, \label{eqn:potdipole}
\eeq
where, for convenience, we have defined the quadratic polynomial
\beq
P(y) &=& y^2(1+\lambda)(\lambda-\nu) \nonumber\\
   &+& y\left[ \lambda^2(3+\nu)+2\lambda(1-3\nu)-(1-\nu) \right] \nonumber\\
   &+& \left[ \lambda^2(4-\nu)-\lambda(3+\nu)+1 \right]\,.
\eeq

Finding the critical value of the angular momentum such that geodesics with $L<L_{\rm crit}$ are captured by the BH (and bounce back to infinity otherwise) is equivalent to requiring the existence of  degenerate roots of the potential $V_y$. Being the expression for the potential cubic in $y$, this calculation normally requires a numerical approach.
However, we will consider below specific cases where simplifications occur and an analytical approach is therefore conceivable.

%%%%%%%%%%%%%%%%%%%%%%%%%%%%%%%%%%%%%%%%%%%%%%%%%%%%%%%%%%%
\subsection{The neutral black ring}
\label{sec:neutrings}
\begin{figure}[htpb!]
\includegraphics[width=8.5cm]{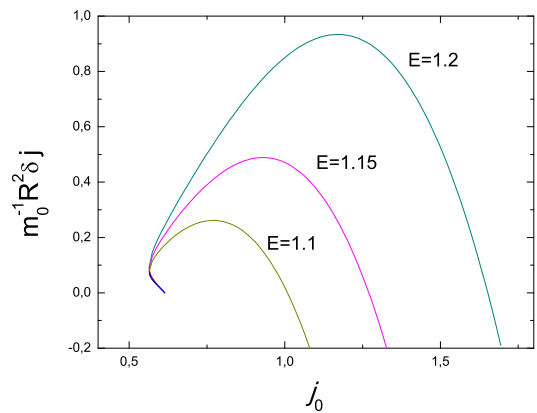}
\caption{\label{fig:FatRingSpin} This figure shows the maximum increase in spin, $\frac{R^2}{m_0}(\delta j)_{\rm max}$ caused by a massive particle with energies per unit mass $E=1.1, 1.15, 1.2$ falling into a singly spinning neutral black ring. Notice that it is not possible to overspin a fat black ring. The lower branch corresponds to the `fat' black rings and the vertical line marks the minimum spin that (regular) black rings can possess.} 
\end{figure}

\begin{figure}[htpb!]
\includegraphics[width=8.5cm]{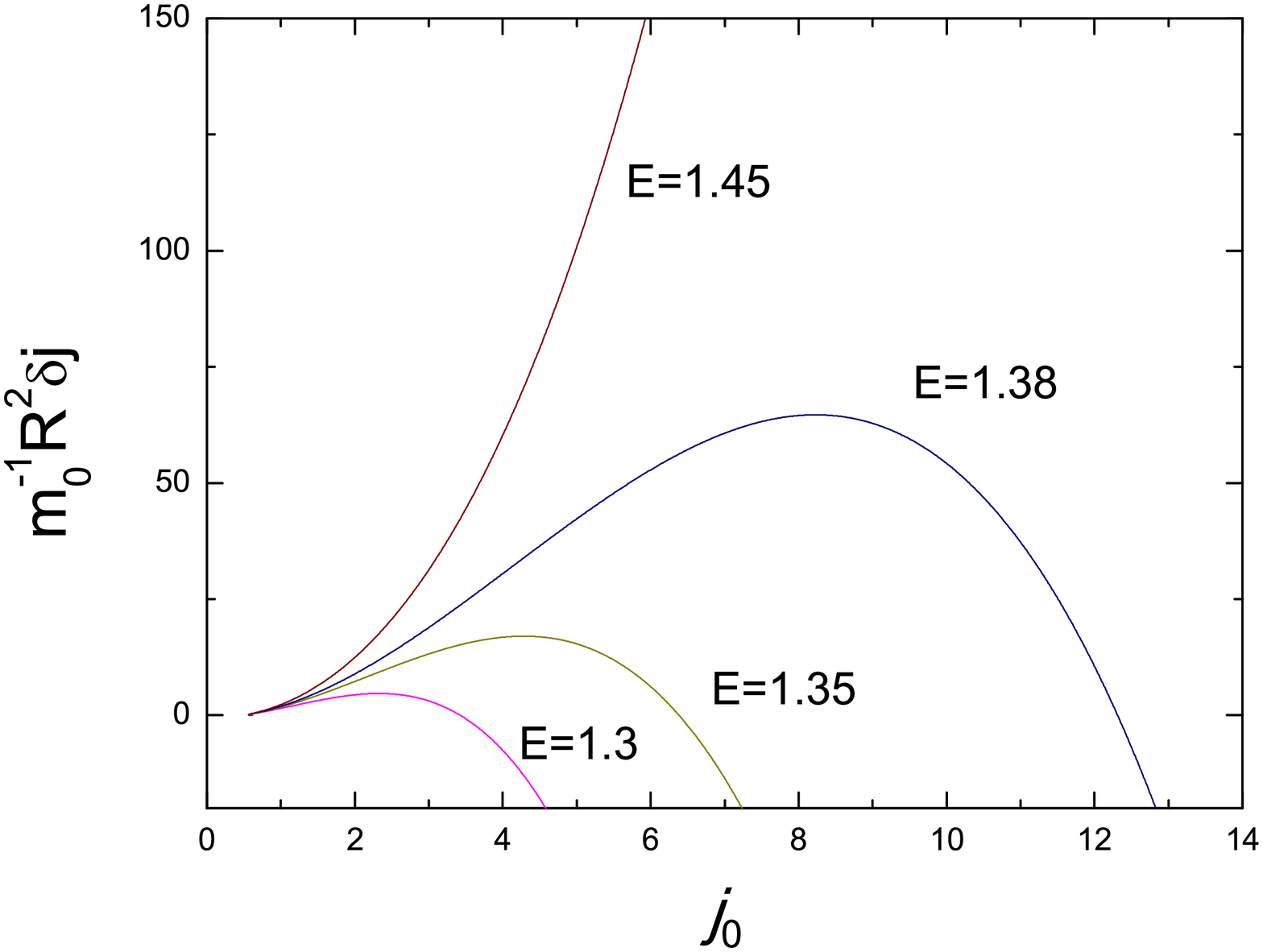}
\caption{\label{fig:ThinRingSpin} This figure shows the maximum increase in spin, $\frac{R^2}{m_0}(\delta j)_{\rm max}$ caused by a massive particle with energies per unit mass $E=1.3, 1.35, 1.38, 1.45$ falling into a singly spinning neutral black ring. The fat black ring branch is not visible in this plot. Thin black rings with very large spins $j_0$ always loose dimensionless angular momentum when absorbing a massive particle with $E<\sqrt{2}$.} 
\end{figure}

%%%%%%%%%%%%%%%%%%%%%%%%%%%%%%%%%%%%%%%%%%%%%%%%%%%%%%%%%%%
For dipole charge parameter $\mu=0$ we recover the neutral black ring solution. It is possible
to see from Eq.~(\ref{eqn:potdipole}) that the potential becomes in this case 
%
%\begin{eqnarray}
%\dot y^2 &=& V_y\,, \nonumber\\
%\frac{R^2 V_y}{(1+y)^3} &=& -\left(\mathcal{C}_2 
%+\frac{\nu\delta_1}{1-\lambda}\right) y^2 \, \nonumber \\
%&+& \left[\mathcal{C}_1+\frac{(1-\nu)\delta_1}{1-\lambda}\right] y
%+ \mathcal{C}_0-\frac{\delta_1}{1-\lambda} \,,
%\label{eqn:potneutral}
%\end{eqnarray}
%
\beq
\dot y^2 &=& V_y\,, \nonumber\\
R^2 V_y &=& (1+y)^3\left[ -\frac{P(y)}{(1-\lambda)^3}E^2 \right. \nonumber\\
  &+& \frac{2C(1+y)^2}{(1-\lambda)^2} E\frac{L}{R} - \frac{(1+y)F(y)}{(1-\lambda)^2} \left(\frac{L}{R}\right)^2 \nonumber\\
  &+& \left. \frac{(1-y)(1+\nu y)}{1-\lambda}\delta_1 \right]\,.
\eeq
By setting the discriminant of the second order equation in $y$ equal to zero we find the equation
for $L_{\rm crit}$, whose solution is given by
\begin{flalign}
&L_{\rm crit}=\frac{R}{\sqrt{1-\nu}} \left[ 2E\sqrt{\nu} \right. \nonumber\\
&+ \left.\sqrt{(E^2-1)(1+\nu)\left(1+3\nu-2\sqrt{2\nu(1+\nu)}\right)} \right]\,.
\label{Lcrit4ring}
\end{flalign}
Together with the expressions for the mass and angular momentum of the singly spinning balanced black ring given in Eq.~(\ref{eqn:dipolemass}) and (\ref{eqn:dipoleJ}) (in the limit $\mu=0$), 
formula~(\ref{Lcrit4ring}) can be inserted into equation~(\ref{evspin}) to yield the maximum addition of angular momentum obtained by throwing a massive particle into the black ring.  Also note that, for the neutral ring,
\be
j_0=\sqrt{\frac{4(1+\nu)^3}{27\pi\nu}}\,. 
\ee
The results are presented in Figs.~\ref{fig:FatRingSpin} and~\ref{fig:ThinRingSpin}.
The thin ring branch ($0<\nu<1/2$) is visible in both figures whereas the fat ring branch ($1/2<\nu<1$) is only apparent in Fig.~\ref{fig:FatRingSpin}~\footnote{One would need to zoom in on $j_0=\sqrt{32/27\pi}$ to see the fat ring branch in Fig.~\ref{fig:ThinRingSpin}.}.  We observe that it is possible to spin-up black rings in the fat branch but the maximum increase in angular momentum vanishes in the singular limit $\nu \rightarrow 1$. This can be shown by using Eqs.~(\ref{Lcrit4ring}) and~(\ref{evspin}) and is in accordance with the results of~\cite{Elvang:2006dd}.  Therefore, fat black rings cannot be over-spun.  Another interesting feature is that, for sufficiently low particle energies (more specifically, $E<\sqrt{2}$), black rings with large spins always see their angular momentum reduced if they absorb the particle.

%%%%%%%%%%%%%%%%%%%%%%%%%%%%%%%%%%%%%%%%%%%%%%%%%%%%%%%%%%%%%%%%%%%%%%%%%%%%
\subsection{The dipole black ring}
\label{sec:diprings}
%%%%%%%%%%%%%%%%%%%%%%%%%%%%%%%%%%%%%%%%%%%%%%%%%%%%%%%%%%%%%%%%%%%%%%%%%%%%
We will now consider the more general case of dipole black rings. As for the neutral ring, these have two branches -- `fat'
and `thin'-- governed by the parameter $\mu$, and the extremal limit
$\nu\rightarrow0$ provides an upper bound for both of them. It is therefore interesting to study
what happens in this specific limit.

\begin{figure}[htpb!]
\includegraphics[width=8.5cm, height=7cm]{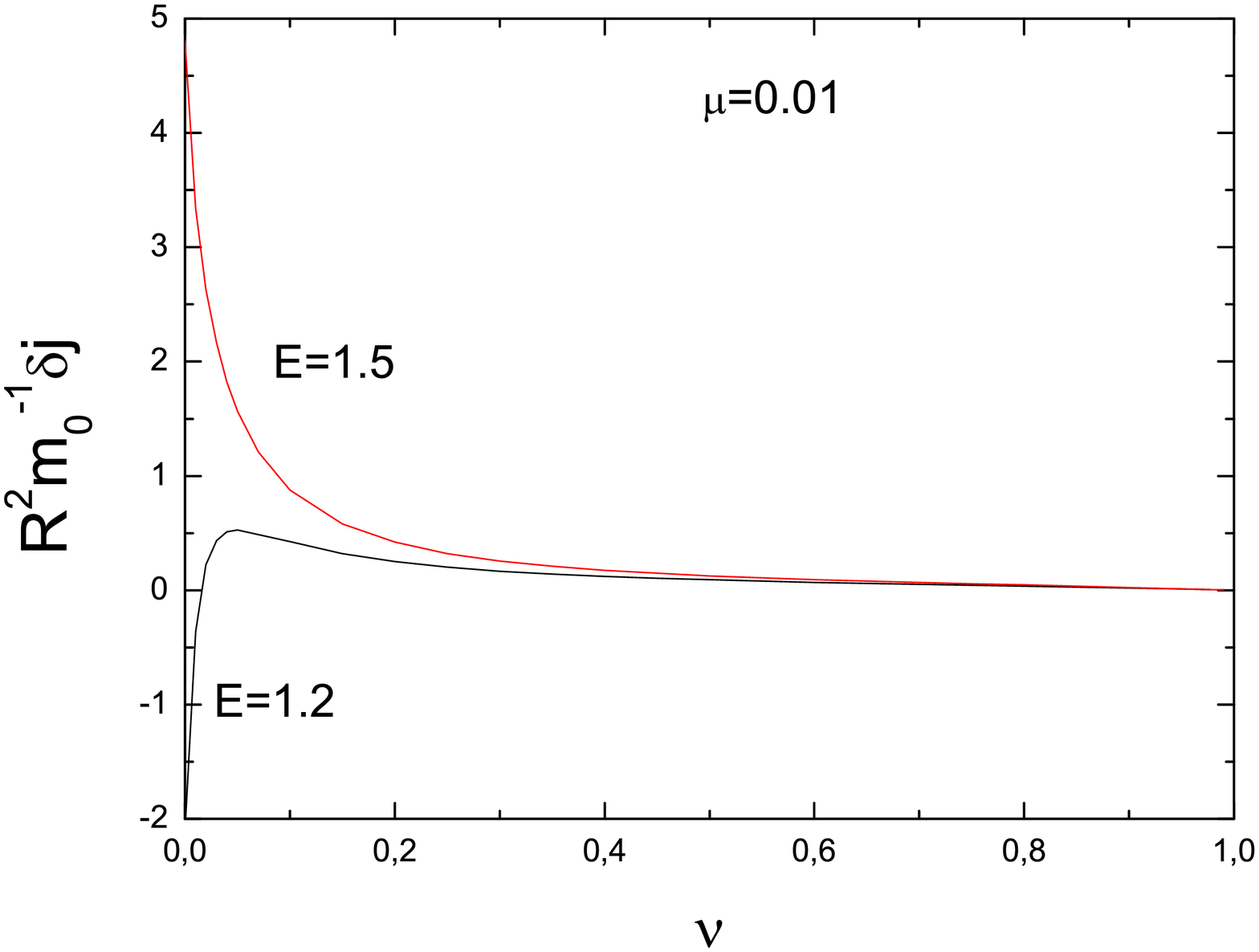}
\caption{\label{fig:Dipolering} This figure shows the maximum increase in spin, $\frac{R^2}{m_0}(\delta j)_{\rm max}$ caused by a massive particle with energies per unit mass $E=1.2$ and $E=1.5$ falling into a dipole black ring. The dipole parameter is set to $\mu=0.01$ while in the horizontal axis the parameter $\nu$, which spans the interval [0,1), is shown.} 
\end{figure}

From the expression of the potential in Eq.~(\ref{eqn:potdipole}) it is possible to solve numerically for the
value of $L_{\rm crit}$, which can be inserted into equation~(\ref{evspin}) to obtain the maximum addition of angular momentum of a massive particle into the dipole black ring. We show the result in Fig.~\ref{fig:Dipolering} for a particle of energies $E=1.2,1.5$ and dipole $\mu=0.01$. 

Let us consider more in detail the extremal limit $\nu\to0$. This limit is hard to tackle and so we consider throwing in massless particles. This amounts to a simplification because when $\delta_1=0$, finding the turning points in the radial potential becomes a quadratic equation:
\beq
R^2 V_y &=& \frac{(1+y)^3}{(1+\mu)H(y)} \left[ -\frac{P(y)}{(1+\mu)(1-\lambda)^3}E^2 \right. \\
  &+& \left. \frac{2C(1+y)^2}{(1+\mu)(1-\lambda)^2} E\frac{L}{R} - \frac{(1+y)F(y)}{(1+\mu)(1-\lambda)^2} \left(\frac{L}{R}\right)^2\right] \nonumber
\eeq
Thus, we can obtain explicitly an expression for the critical angular momentum $L_{\rm crit}$. Nevertheless, the formula is intractable and so we proceed in the manner we now describe.

\begin{figure}[htpb!]
\includegraphics[width=8.5cm, height=7cm]{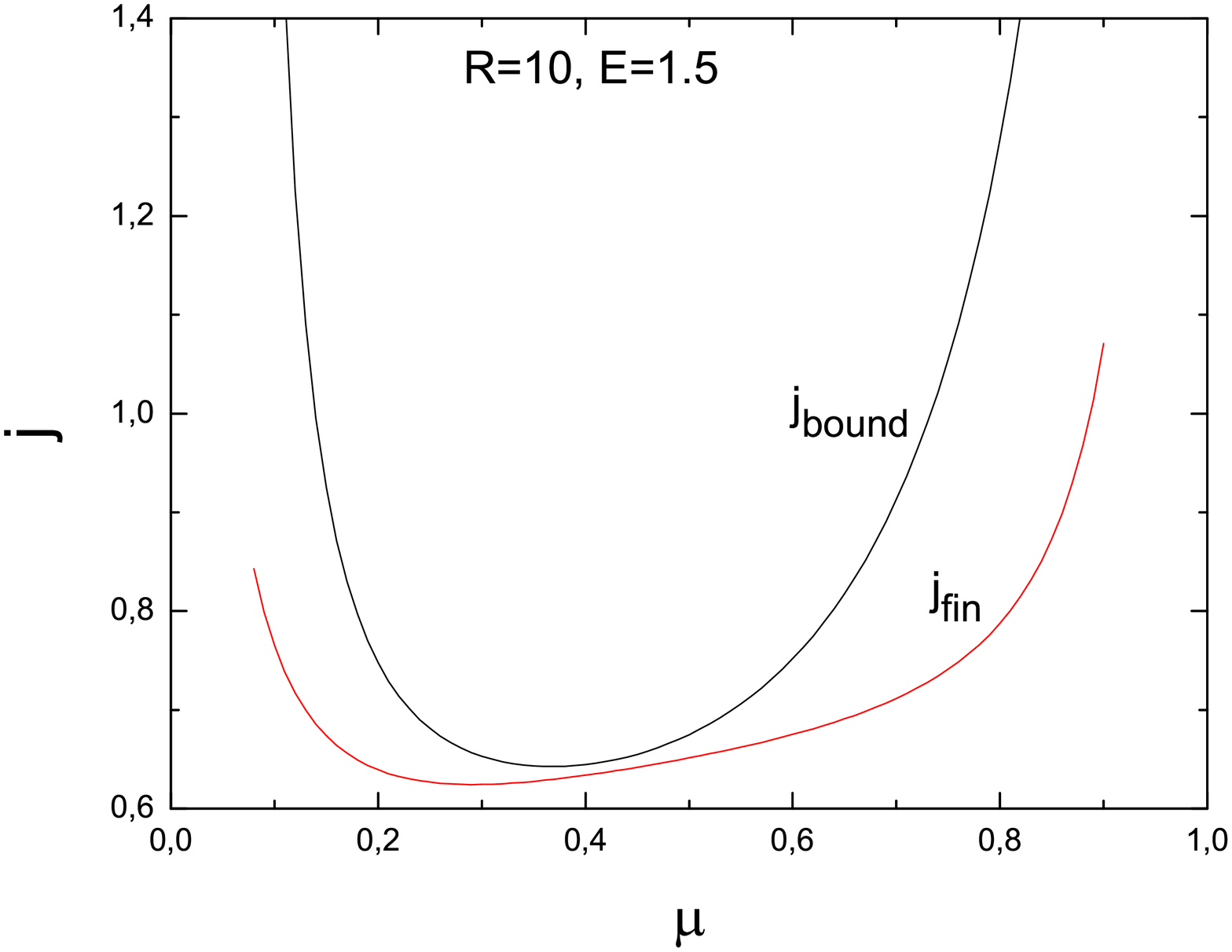}
\caption{\label{fig:DipoleExtremal} This figure shows the final dimensionless angular momentum, $j_{\rm fin}$, caused by a massless particle with energy $E=1.5$ falling into an extremal dipole black ring. The horizontal axis corresponds to the dipole parameter $\mu$ of the initial ring. The value $R=10$ was employed to ensure that $E\ll {\cal M}_0$ is always satisfied, therefore avoiding backreaction effects. Also shown is the upper (extremal) bound on $j$ considering the final ring has dipole charge $q_{\rm fin}$. It is apparent that the dipole ring is never overspun.} 
\end{figure}

Assume an initial dipole ring already at extremality and with some dipole parameter $\mu$. One can then easily compute the initial quantities $j_0$ and $q_0$. Next we determine $(\delta j)_{\rm max}$ and $\delta q$ using Eqs.~(\ref{evspin}) and the expression for $\delta q$ given by 

\be
\delta q = -\frac{m_0}{{\cal M}_0} \frac{E q_0}{2}\,,
\label{deltaq}
\ee
setting $m_0=1$ in these expressions since we are now considering massless particles. After absorption of this particle the black ring will be characterized by the quantities
\beq
j_{\rm fin} &=& j_0 + (\delta j)_{\rm max}\,, \\
q_{\rm fin} &=& q_0 + \delta q\,.
\eeq
But for the final dimensionless dipole charge thus obtained, $q_{\rm fin}$, we can compute the corresponding upper bound on the dimensionless angular momentum, $j_{\rm bound}$. These results are presented in Fig.~\ref{fig:DipoleExtremal} for a range of initial dipole parameter $\mu$. Also shown is the upper (extremal) bound on $j$ considering the final ring has dipole charge $q_{\rm fin}$. We find that $j_{\rm fin}<j_{\rm bound}$, independently of $\mu$. This provides clear indication that the dipole ring cannot be spun above extremality.

%%%%%%%%%%%%%%%%%%%%%%%%%%%%%%%%%%%%%%%%%%%%%%%%%%%%%%%%%%%%%%%%%%%%%%%%%%%%%%%%
\section{Discussion}
%%%%%%%%%%%%%%%%%%%%%%%%%%%%%%%%%%%%%%%%%%%%%%%%%%%%%%%%%%%%%%%%%%%%%%%%%%%%%%%%
We have shown that several different black hole geometries are immune to the throwing of
point particles: in the geodesic approximation employed here, particles which are
captured by the black hole have an angular momentum which is sufficiently low so as to be harmless; in
fact sufficiently low that they are never able to spin-up the geometry past the extremal
value. It seems unlikely that taking radiation reaction into account will alter these
conclusions. Our results should be taken together with full-blown numerical evolutions in
four-dimensional spacetimes, where it was shown that the collision of equal-mass black
holes at generic velocities never produces a naked singularity \cite{Sperhake:2009jz}.
It is thus tempting to conjecture that this is a general result, and that black
hole-black hole collisions at arbitrary velocity are governed by some kind of Cosmic
Censor.

We have only dealt with asymptotically flat spacetimes. It would surely be interesting to
generalize the present results to say, (anti-)de Sitter backgrounds. More interesting yet
would be to understand how do geodesics
convey information about the event horizon: clearly the maximum impact parameter for
capture ``conspires'' with the properties of the black hole in such a way as to never allow
singularities to form. Is this really just a coincidence or is it forced on us by the
field equations?
Whatever the answer, Cosmic Censorhip remains a fascinating topic in gravitation.
%%%%%%%%%%%%%%%%%%%%%%%%%%%%%%%%%%%%%%%%%%%%%%%%%%%%%%%%%%%%%%%%%%%%%%%%%%%%%%%%%%%%%%%%%%%%%%%%%%%%%%%%%%%%
\section*{Acknowledgements}
%%%%%%%%%%%%%%%%%%%%%%%%%%%%%%%%%%%%%%%%%%%%%%%%%%%%%%%%%%%%%%%%%%%%%%%%%%%%%%%%%%%%%%%%%%%%%%%%%%%%%%%%%%%%%

The authors would like to thank Roberto Emparan for clarifying correspondence.
V.C. is supported by a ``Ci\^encia 2007'' research
contract and by Funda\c c\~ao Calouste Gulbenkian through a short-term
scholarship. MBL, AN and  JVR acknowledge financial support from {\it Funda\c{c}\~ao para a Ci\^encia e Tecnologia} (FCT)-Portugal through the fellowships SFRH/BPD/26542/2006, SFRH/BPD/47955/2008 and SFRH/BPD/47332/2008, respectively. This work was partially funded by {\it Funda\c c\~ao para a Ci\^encia e Tecnologia} (FCT)-Portugal through projects
PTDC/FIS/64175/2006, PTDC/ FIS/098025/2008, PTDC/FIS/098032/2008 and CERN/FP/109290/2009.
The authors thankfully acknowledge the computer resources, technical expertise and
assistance provided by the Barcelona Supercomputing Center - Centro Nacional
de Supercomputaci\'on.

%%%%%%%%%%%%%%%%%%%%%%%%%%%%%%%%%%%%%%%%%%%%%%%%%%%%%%%%%%%%%%%%%%%%%%%%%%%%%%%%

\end{document}